# Deposition of $Ga_2O_3$ thin films by liquid metal target sputtering

Martins Zubkins[1*], Viktors Vibornijs[1], Edvards Strods[1], Edgars Butanovs[1], Liga Bikse[1], Mikael Ottosson[2], Anders Hallén[3], Jevgenijs Gabrusenoks[4], Juris Purans[1], Andris Azens[4]

[1]Institute of Solid State Physics, University of Latvia, Kengaraga 8, LV-1063 Riga, Latvia

[2]Department of Chemistry, Ångström Laboratory, Lägerhyddsvägen 1, SE 75120 Uppsala, Sweden

[3]KTH Royal Institute of Technology, School of EECS, P.O. Box Electrum 229, SE 16440, Kista-Stockholm, Sweden

[4]AGL Technologies SIA, Smerla 3, LV-1006 Riga, Latvia

*Corresponding author: martins.zubkins@cfi.lu.lv

## Abstract

This paper reports on the deposition of amorphous and crystalline thin films of $Ga_2O_3$ by reactive pulsed direct current magnetron sputtering from a liquid gallium target onto fused (f-) quartz and *c* plane (c-) sapphire substrates, where the temperature of the substrate is varied from room temperature (RT) to 800°C. The deposition rate (up to 37 nm/min at RT on f-quartz and 5 nm/min at 800°C on c-sapphire) is two to five times higher than the data given in the literature for radio frequency sputtering. Deposited onto unheated substrates, the films are X-ray amorphous. Well-defined X-ray diffraction peaks of β-$Ga_2O_3$ start to appear at a substrate temperature of 500°C. Films grown on c-sapphire at temperatures above 600°C are epitaxial. However, the high rocking curve full width at half maximum values of ≈ 2.4–2.5° are indicative of the presence of defects. A dense and void-free microstructure is observed in electron microscopy images. Composition analysis show stoichiometry close to $Ga_2O_3$ and no traces of impurities. The optical properties of low absorptance (<1%) in the visible range and an optical band gap of approximately 5 eV are consistent with the data in the literature for $Ga_2O_3$ films produced by other deposition methods.

Key words: gallium oxide, thin films, magnetron sputtering, liquid metal target

**Introduction**

Gallium oxide ($Ga_2O_3$) has attracted a lot of attention as an ultra-wide bandgap semiconductor [1,2]. It can crystallize in five different phases, of which the most thermodynamically stable and technologically relevant is the monoclinic β phase [3,4], although there is also increasing interest in the other phases [5–7] and amorphous coatings [8–10]. The properties of greatest technological importance are transparency in the UV part of the spectrum due to its wide band gap of approximately 4.9 eV, high carrier mobility (up to 200 $cm^2$/Vs), high breakdown field (8 MV/cm), and high thermal and chemical stability [1]. Practical applications include solar-blind photodetectors, high-voltage transistors, high-power Schottky diodes, high-temperature chemical sensors, and transparent electrical conductors in optoelectronic devices [11–15].

The known methods of $Ga_2O_3$ thin film deposition include metal-organic chemical vapor deposition [16], plasma-enhanced chemical vapor deposition [17], molecular beam epitaxy [18], atomic layer deposition [19], pulsed laser deposition [20], e-beam evaporation [9], and radio-frequency (RF) magnetron sputtering (MS) from ceramic $Ga_2O_3$ targets [21]. While MS is generally known for high-quality coatings produced under well controlled and reproducible conditions [22–24], the drawback of the RF mode of sputtering is its impractically low sputtering rate. Moreover, this technique has scalability issues, as RF power supplies have a limited maximum power (on the order of a few kW).

Although the usual source of the sputtered particles is a solid target, it has been demonstrated that gallium nitride films can also be deposited by reactive MS from a liquid gallium target [25]. In this study, we used reactive pulsed DCMS from a liquid gallium target for the deposition of $Ga_2O_3$ thin films. The goals of the study were twofold: (i) to establish whether a robust sputtering process could be set up and optimized with high deposition rates compared to RF sputtering from ceramic targets, and (ii) to verify that the film composition, structure and optical properties were those of stoichiometric $Ga_2O_3$. We found that highly transparent amorphous and crystalline β-$Ga_2O_3$ coatings could be created at deposition rates two to five times higher than those typical for RF sputtering.

**Experimental details**

$Ga_2O_3$ thin films were deposited by reactive pulsed DCMS from a gallium metal target in an Ar/$O_2$ atmosphere. As the melting temperature of gallium is 29.8°C, the target was in a liquid state during sputtering. To prevent the melted metal from flowing off the magnetron surface, a box-shaped stainless steel target container was used with 3-mm-high walls and a machined recess in the base plane (Fig. 1(a) and Fig. S1 in the supplementary material). To prevent the liquid gallium from contracting into islands and leaving parts of the container base plate exposed to sputtering, the container surface was pre-coated with a wettability-promoting layer of carbon. The target was prepared by melting metallic gallium pellets (PI-KEM, 99.999%) and then cooling the container to allow the target to solidify prior to being placed onto the magnetron.

Film deposition was performed using a G500M.2 PVD coater equipped with a planar balanced magnetron (Sidrabe Vacuum, Ltd.). The magnetron was placed under the substrate holder at a distance of 11 cm. The same cooling system was used as for ordinary solid targets, i.e., the magnetron was cooled by a flow of water (≈2.5 l/min) at 20°C. The base pressure was ≤7×10$^{-4}$ Pa in a turbomolecular pumped chamber. The residual gas composition at the base pressure of 6.4×10$^{-4}$ Pa is shown in Fig. S2, and consisted of approximately 92% (5.9×10$^{-4}$ Pa) $H_2O$, 2% (1.3×10$^{-5}$ Pa) $N_2$, 2% (1.2×10$^{-5}$ Pa) $CO_2$, 1% (1.0×10$^{-5}$ Pa) $O_2$, 1% (7.0×10$^{-6}$ Pa) $H_2$, and total contamination of other gases <2%. The process pressure was set to 0.4 Pa by feeding 30 sccm (standard cubic centimeters per minute) of argon and partly closing the throttle valve between the chamber and the pump. Sputtering was carried out in pulsed DC (80 kHz, $t_{rev}$ = 2.5 μs) mode in an $O_2$ (99.999% pure)/Ar (99.9999% pure) atmosphere at a power of 150 W (1.3 W/cm$^2$). The amount of residual gases in the sputtering atmosphere was less than 0.3%. Fused (f-) quartz (SPI Supplies) and c-plane (c-) sapphire (Biotain Crystal) substrates were used, and the substrate temperature was varied between room temperature (RT) (i.e., no intentional heating provided) and 800°C. A few selected samples deposited at RT and 500°C were post-annealed for 5 h in air at 800°C and 700°C, respectively, to determine whether there were differences in the properties of films heated during or after deposition.

Control of the sputtering process was achieved by optical emission spectroscopy (OES), i.e., the oxygen flow was controlled by a feedback loop keeping the ratio $I_{proc}/I_{met}$ constant, where $I_{proc}$ and $I_{met}$ are the intensities of the gallium 417.2 nm emission line in the process and metal modes, respectively. An example of a typical deposition cycle is shown in Fig. 1(b). The process was started in the metal mode (feeding argon only) with the shutter closed in front of the substrate. The target was conditioned until $I_{met}$ became constant. Oxygen was added and the flow was increased manually to reduce the intensity of the emission line to the desired value, $I_{proc}$. The process was locked into PID (proportional-integral-derivative) control mode, to keep the $I_{proc}/I_{met}$ ratio constant. The shutter was opened, and the film was deposited onto the substrate. In the specific example shown in Fig. 1(b), the oxygen flow during the deposition was approximately 7.8 sccm. The reason for using the ratio rather than $I_{proc}$ was to achieve better process reproducibility over a long period of time. While a few consecutive deposition cycles could be successfully carried out by using the absolute value of $I_{proc}$, taking into account $I_{met}$ compensates for any changes in the light collection efficiency due to instrumental factors, for example the protective window of the optical channel gradually becoming coated over the course of many cycles, or minor changes in the interior geometry of the chamber introduced by service and maintenance. This makes the ratio $I_{proc}/I_{met}$ a more robust parameter than $I_{proc}$ alone.

The film thickness was measured using a stylus profilometer (Veeco Dektak 150) and ellipsometry (Woollam RC2-XI), and both techniques gave similar values for the thickness. The thickness of all films was in the range 60–780 nm.

The crystallographic structure of the films was examined by X-ray diffraction (XRD), using a Rigaku MiniFlex600 X-ray powder diffractometer with Bragg-Brentano θ-2θ geometry and a 600 W Cu anode (Cu Kα radiation, λ = 1.5406 Å) X-ray tube. The pole figures and the rocking curves were measured with a Philips MRD Pro Diffractometer. The system was set up with copper Cu Kα radiation using point focus and a capillary lens, with a beam size of 2 × 2 mm. A 0.18 parallel plate collimator and graphite monochromator were used as primary and secondary optics, respectively. The pole figures for the (–401) plane were measured for PSI = 0–80°.

The optical transmittance and reflectance of the films in the range 200–2000 nm were measured by an Agilent Cary 7000 spectrophotometer. The sample was placed at an angle of 6° against the incident beam of P-polarized light, and the detector was placed at an angle of 180° behind the sample to measure the transmittance and at 12° in front of the sample to measure the specular reflectance.

X-ray photoelectron spectrometry (XPS) measurements were performed by an ESCALAB 250Xi (ThermoFisher) instrument to determine the film composition. An Al Kα X-ray tube with energy 1486 eV was used as an excitation source, the size of the analyzed sample area was 650 × 100 μm, and the angle between the analyzer and the sample surface was 90°. An electron gun was used to perform charge compensation. The base pressure during spectra acquisition was better than $10^{-5}$ Pa. Prior to analysis, the samples were sputter-cleaned for 60 s with an argon gas cluster ion beam (atom cluster size 150, energy 4000 eV, area 2 × 2 mm) at an incidence angle of 30° with respect to the normal of the surface.

Elastic recoil detection analysis (ERDA) utilizing 36 MeV $^{127}$I incident ions and time-of-flight detection was performed to determine the chemical purity of the films [26].

The surfaces of the films were characterized by scanning electron microscopy (SEM), using a Thermo Scientific Helios 5 UX dual-beam microscope. Cross-section studies were performed using a transmission electron microscope (TEM, Tecnai G20, FEI) operating at 200 kV. Lamellae were prepared by a focused ion beam (FIB) technique. To protect the films during preparation of the lamellae, the surface was coated with a 30-nm-thick Au layer and a 1.5-μm-thick Pt layer before FIB exposure.

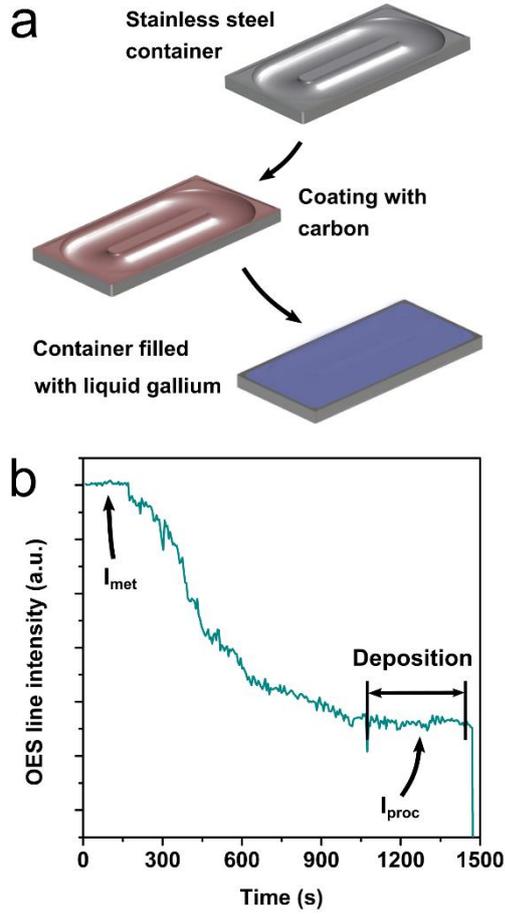

**Fig. 1**. (a) Schematic diagram of the target preparation; (b) example of control of the deposition process by optical emission spectroscopy, where the intensity of the gallium 417.2 nm emission line was controlled by a PID loop to adjust the oxygen flow. The cycle was initiated by sputtering in argon (line intensity $I_{met}$). Oxygen was added and the flow increased until the process line intensity $I_{proc}$ reached the $I_{proc}/I_{met}$ level chosen for deposition, and was then kept at $I_{proc}/I_{met}$ by the PID for film deposition onto the substrate.

**Results and discussion**

The deposition rates as function of $I_{proc}/I_{met}$ for the films deposited on f-quartz and c-sapphire for varying substrate temperatures are shown in Fig. 2, where larger values of $I_{proc}/I_{met}$ correspond to less oxygen in the sputtering atmosphere. For the data points shown in Fig. 2, the films deposited at rates of up to 37 nm/min ($I_{proc}/I_{met}$ = 0.69) were highly transparent. Higher $I_{proc}/I_{met}$ values (and hence higher rates) yielded apparently underoxidized (optically absorbing) films.

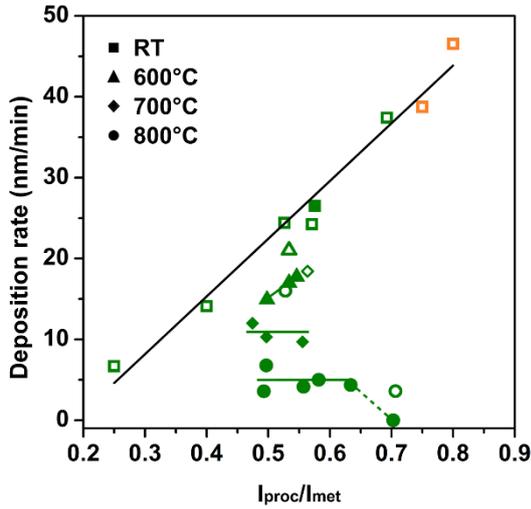

**Fig. 2.** Deposition rate of $Ga_2O_3$ film as a function of normalized Ga 417.2 nm emission line intensity ($I_{proc}/I_{met}$) for OES-controlled sputtering from metallic gallium target in $Ar/O_2$ atmosphere for substrates at room temperature (squares), 600°C (triangles), 700°C (diamonds), 800°C (circles) onto f-quartz (open symbols) or c-sapphire (solid symbols) substrates. Larger values of $I_{proc}/I_{met}$ correspond to less oxygen in the sputtering atmosphere. The lines between the deposition rate points on quartz and sapphire substrates at room temperature (black line) and on sapphire substrates at elevated temperatures (green lines) are drawn to guide the eye. The orange symbols represent underoxidized (optically absorbing) films.

The main takeaways from Fig. 2 are as follows:

(i) For unheated substrates, the trend of the deposition rate with $I_{proc}/I_{met}$ is what one would expect for a typical oxide coating deposited in a reactive process, i.e., the rate increases upon reducing the oxygen content in the sputtering atmosphere. The linearity between the deposition rate and $I_{proc}/I_{met}$ suggests that (a) the amount of light emitted by the sputtered atoms and detected by OES is directly proportional to the number of sputtered atoms and hence to the number of atoms arriving at the substrate, and (b) the higher order effects that could potentially disrupt the linearity (for instance, the optical excitation yield changing with $I_{proc}/I_{met}$) are neglectable within the studied range of deposition conditions.

(ii) Within the present statistics, there is no difference in the deposition rates for unheated f-quartz and c-sapphire substrates.

(iii) For a given $I_{proc}/I_{met}$, the deposition rate decreases with the substrate temperature, and the decrease is substrate-dependent: while the trend is the same for both substrates, the effect is stronger for c-sapphire than for f-quartz. Since the sputtering conditions are unchanged and the rate of arrival of atoms at the substrate is the same for both substrates at any temperature, the apparent reason for the decrease in deposition rate is the decrease in an atom's probability of sticking to the substrate. The exact reason for the dependence of the rate on the substrate is a matter for future study, but we note that one difference detected in this study is that the films grown on c-sapphire are epitaxial, while the films on f-quartz consist of differently oriented crystallites (see the XRD data below). The remarkably strong effect of the substrate temperature is consistent with the data in the literature for RF sputtering. $Ga_2O_3$ films have been RF sputter-deposited on 800°C c-sapphire substrates at 1 nm/min [27–30], in sharp contrast with the rates of 15–22 nm/min reported for films deposited onto unheated glass substrates [31]. A decrease in the deposition rate from approximately 20 nm/min to less than 10 nm/min by increasing the silicon substrate temperature from 100°C to 600°C was reported in [32].

(iv) The slope of the deposition rate vs. $I_{proc}/I_{met}$ changes with temperature, from its steepest value at a substrate temperature of RT (black line in Fig. 2), to less steep at 600°C, to practically independent of $I_{proc}/I_{met}$ at 700°C and 800°C (green lines in Fig. 2). Again, since the sputtering conditions are unchanged with temperature, the change in slope reflects a shift in the factor limiting the deposition rate. At RT, the deposition rate is limited by the rate of arrival of the atoms at the substrate, and a higher rate of arrival at larger $I_{proc}/I_{met}$ results in a higher deposition rate. As the substrate temperature increases, the increase in the rate of arrival of the atoms does not result in an equally high increase (at 600°C), or any increase at all (at 700°C and 800°C) in the deposition rate with $I_{proc}/I_{met}$, suggesting that the factor limiting the deposition rate is gradually shifting from the rate of arrival to the rate at which the atoms stick to the surface.

(v) There is a sharp drop in the deposition rate to virtually zero once $I_{proc}/I_{met}$ exceeds a certain value (approximately 0.7) for a substrate temperature of 800°C for c-sapphire. Since the sputtering yield, and hence the arrival rate of gallium atoms at the substrate, increases with $I_{proc}/I_{met}$ irrespective of the substrate temperature, this drop in the deposition rate is indicative of a certain minimum oxidation rate required for a film to be formed on the substrate. If the oxidation rate is below this minimum level, the arriving gallium atoms appear to be re-evaporated from the surface at a rate exceeding their rate of arrival.

(vi) The deposition rate achieved for optically transparent films (up to 37 nm/min at RT on f-quartz and 5 nm/min at 800°C on c-sapphire) is higher than the data in the literature for RF sputtering (22 and 1 nm/min, respectively). In addition to the reduced process time, another benefit from the higher deposition rate is the higher purity of the films. The main source of contamination for sputter deposited films (in the context of gallium oxide, this is often referred to as unintentional doping [1]) are the residual gas molecules present in the chamber at the base pressure. For a constant arrival rate of these molecules at the substrate at a given base pressure, the higher arrival rate of gallium atoms means fewer contaminants per incoming gallium atom in the growing film.

XRD patterns for the as-deposited and post-annealed films are shown in Fig. 3. The films deposited on f-quartz and c-sapphire substrates at RT were X-ray amorphous (Fig. 3(a,b)); however, the films on f-quartz showed a broad, relatively weak yet detectable XRD band between 29° and 39° (Fig. S3 in the Supplementary Material). Similar XRD features reported in Refs. [29,31] were attributed to the onset of crystallization of β-$Ga_2O_3$, suggesting that crystallites of very small size may be present even in the predominantly amorphous films deposited at RT.

For the films deposited on f-quartz, clearly detectable crystallization starts between 500°C and 600°C (Fig. 3(a)). The maximum at ≈30.5° can be ascribed to either the (400) or (−401) plane of the β-$Ga_2O_3$ phase, according to ICDD card 01-087-1901. When the deposition temperature is raised to 700°C and above, the maximum becomes slightly asymmetric (with a shoulder towards smaller angles), and other low-intensity maxima (−201) and (002) appear at 19° and 32°, respectively, indicating a slight change in the texture. The origin of the asymmetric shape is most likely a superposition of signals coming from several planes, i.e., (400), (−401), or (110). The X-ray amorphous films deposited at RT can be crystallised by post-annealing (Fig. 3(a)); however, the texture of these films differs from that of those deposited at high (≥ 600°C) temperatures, in that the (002) maximum in the annealed films is more intense and comparable to the maximum at ≈30.5°, while the (−201) maximum is not present.

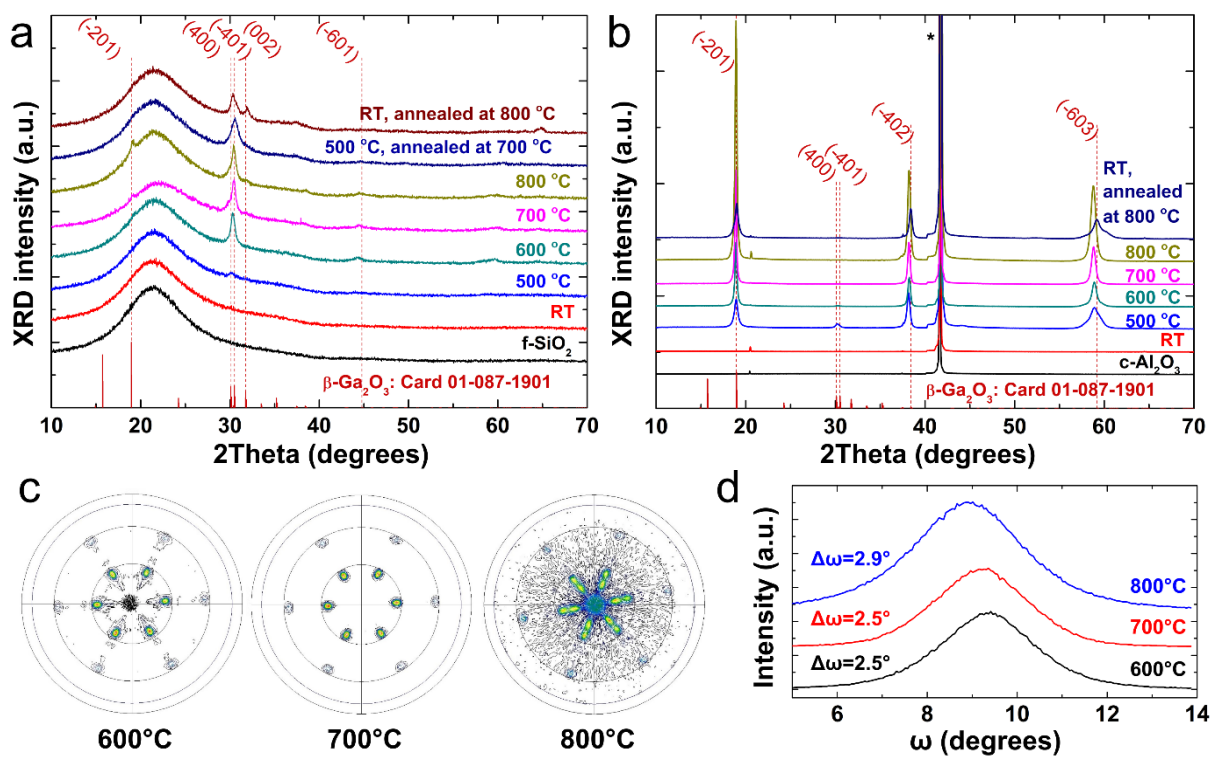

**Fig. 3.** X-ray diffractograms of as-deposited and post-annealed $Ga_2O_3$ films on (a) f-quartz and (b) c-sapphire substrates at different temperatures, including the signals from the substrates. The angles of the Bragg peaks for monoclinic β-$Ga_2O_3$ (ICDD card 01-087-1901) are shown by vertical red lines. The maximum (006) for the c-sapphire substrate is denoted by *. (c) Pole figures for the (−401) plane and (d) ω scans of (–201) reflections are shown for films deposited on c-sapphire at substrate temperatures of 600°C, 700°C, and 800°C. The films were deposited at $I_{proc}/I_{met} \approx 0.5$.

For the films deposited on c-sapphire, the onset of crystallization is similar to those deposited on f-quartz, i.e., clearly detectable XRD maxima for the β-$Ga_2O_3$ phase appear at substrate temperatures of 500°C and higher (Fig. 3(b)). The pole figures in Fig. 3(c) show a six-fold symmetry indicative of epitaxial growth of the films deposited at 600°C, 700°C, and 800°C. However, the rocking curve full width at half maximum (FWHM) values in Fig. 3(d) of approximately 2.4–2.5° suggest that defects and twins are present in the films. No clear correlation was observed between the FWHM values and the deposition temperature.

Surface morphology and cross-sectional images of the crystalline β-$Ga_2O_3$ films obtained by electron microscopy are shown in Fig. 4. The surface of the film deposited on f-quartz at 700°C exhibits grainy features with sharp edges, random shapes, and an average size of less than 100 nm. The surface becomes visibly smoother when the films are grown at 800°C. The films grown on c-sapphire show a smooth and void-free surface morphology, with some flat, randomly shaped features still present in the images. Dense structures are in evidence in the cross-sectional images. The β-$Ga_2O_3$ films prepared at 700°C on both f-quartz and c-sapphire are highly crystalline; however, while the grains in the film on f-quartz are slightly misoriented, a specific crystallographic direction is clearly visible in the film on c-sapphire, indicating an oriented growth. In this case, the interplanar distance was measured to be 4.7 Å, which matches well with the [−201] growth direction of β-$Ga_2O_3$ on a (0001) sapphire substrate [33].

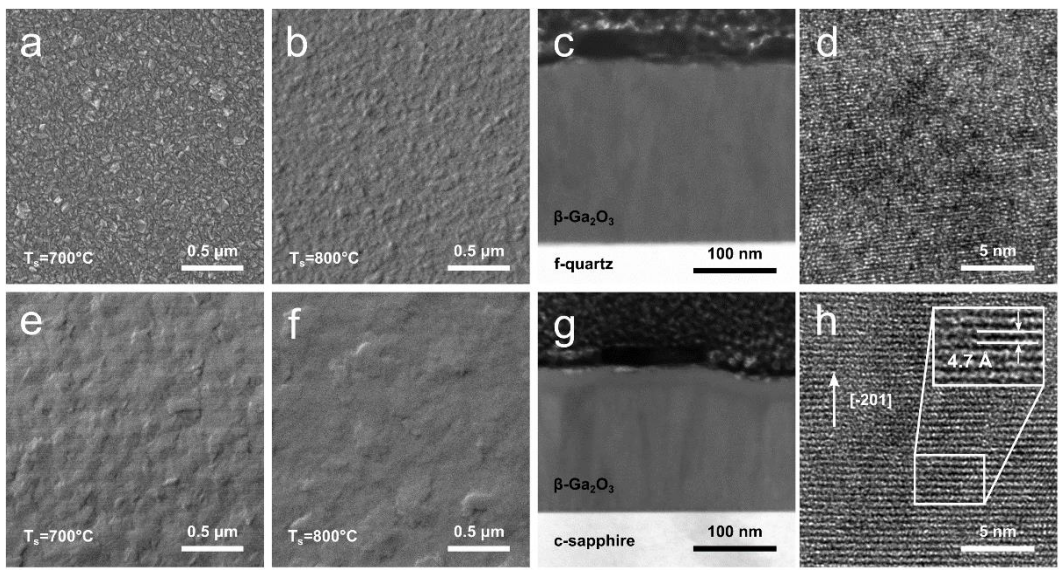

**Fig. 4.** SEM images of the β-$Ga_2O_3$ films on (a, b) f-quartz and (e, f) c-sapphire prepared at (a, e) 700°C and (b, f) 800°C, and of the films' cross-sectional lamellae on (c) f-quartz and (g) c-sapphire prepared

at 700°C. HR-TEM images of the cross-sections (lamellae) of the β-$Ga_2O_3$ films, prepared at 700°C, on (d) f-quartz and (h) c-sapphire. The insets in (h) show the direction of growth and the measured lattice spacing. The films were deposited at $I_{proc}/I_{met} \approx 0.5$.

The composition of the films was verified with XPS and ERDA measurements. A survey scan of $Ga_2O_3$ film deposited on f-quartz at RT (Fig. 5(a)) indicated the presence of Ga and O only, with the Ga/O atom ratio closely matching that of the $Ga_2O_3$ compound, and no peaks of possible contaminants were detected. Similar XPS results were also obtained for films prepared at higher temperatures on the f-quartz and c-sapphire substrates. Fig. 5(b) shows the ERDA results for a sample grown on c-sapphire at 700 °C and $I_{proc}/I_{met} \approx 0.5$. The figure displays the counts for coincidently detected particles at the energy (*x*-axis) and time-of-flight (*y*-axis) detectors. For practical reasons, the *y*-axis is inverted, meaning that ions with low mass, such as oxygen, will have the longest flight times in the ERDA plots. The particles ejected from the surface of the film will have the highest energy, while particles coming from deeper layers will have lower energy, due to the energy loss in passing through the film, and the energy of the incident primary iodine ions will also be reduced, since they had to penetrate deeper into the film. Since the energy is proportional to the square of the inverse flight time, plotting the data in this manner produces square, root-shaped traces for the different elements. The particles emanating from the surface are seen in the upper right-hand side of each trace. Only four elements are shown in the figure: oxygen, aluminum, gallium and iodine. The iodine counts build up in the spectra during the measurement, since previously incident iodine projectiles become recoils. The iodine therefore does not originate from the as-deposited films. Both O and Ga are present at the surface, while the Al trace starts deeper in the sample, since it originates from the substrate. The ratio Ga/O suggests a 1–2 atomic% oxygen deficiency in the film, although this is within the range of potential calibration errors. The film thickness is estimated at about 200 nm, which is in good agreement with the results from the profilometer.

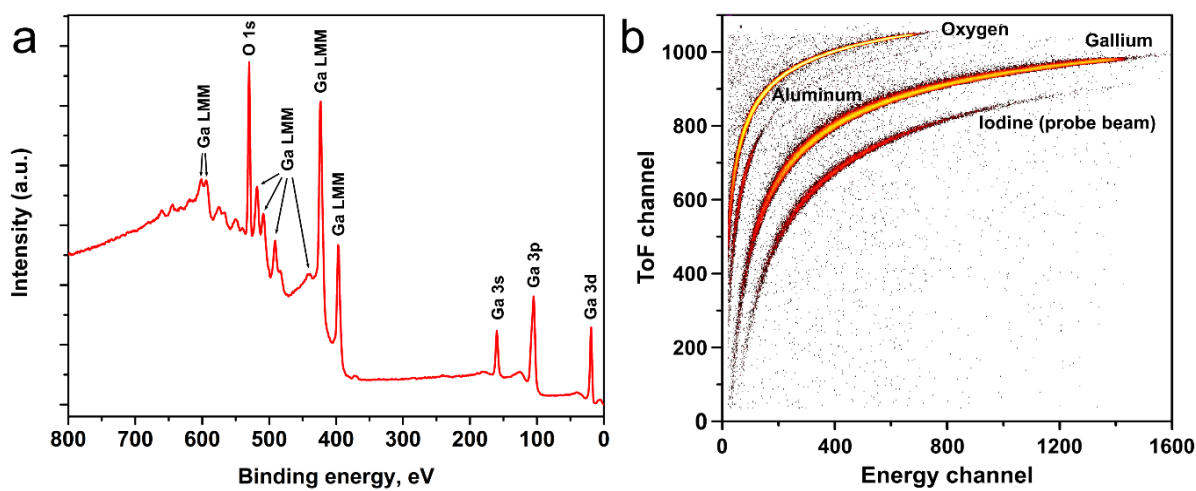

**Fig. 5**. (a) XPS survey scan of $Ga_2O_3$ film deposited on f-quartz at room temperature, showing the elemental composition; (b) ERDA spectrum of β-$Ga_2O_3$ films deposited on c-sapphire at 700°C for $I_{proc}/I_{met} = 0.5$.

Fig. 6(a) shows the transmittance and specular reflectance spectra of selected samples, as-deposited $Ga_2O_3$ films on f-quartz and c-sapphire substrates, for wavelengths from 200 to 2000 nm. The average transmittance is from 80% to 85%, and is mainly limited by the reflectance of ~15–20%. The low amplitude of the interference fringes for the films on c-sapphire is most likely to be due to the apparently similar refractive indices of the film and the substrate. The absorptance spectra of the films in the narrow wavelength range (200–350 nm) can be seen in Fig. 6(b). A slight blue-shift with the deposition temperature is observed for both types of substrates. The absorptance spectra of other films produced in this study can be found in Figs. S4 and S5. The absorptance in the visible range stays low (< 1%) and is typical for stoichiometric $Ga_2O_3$ films for all samples deposited at $I_{proc}/I_{met} \leq 0.7$. An attempt to increase the deposition rate (more specifically, depositing the films at $I_{proc}/I_{met} > 0.7$ and RT) yielded a sub-stoichiometric $Ga_2O_{3-x}$ film with increased absorptance (Fig. S4).

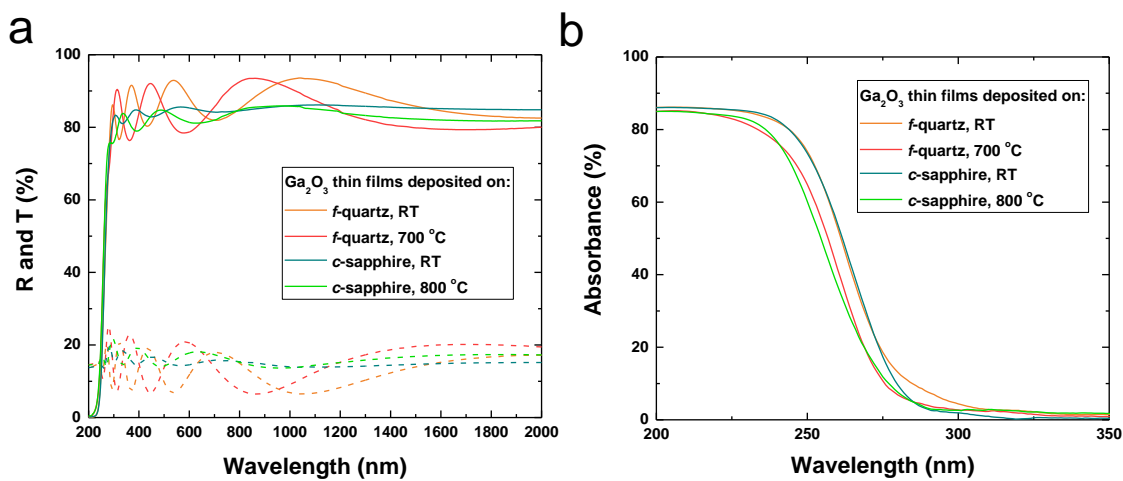

**Fig. 6.** (a) Transmittance (T) and specular reflectance (R) spectra (solid and dashed lines, respectively) in the range 200–2000 nm for $Ga_2O_3$ films deposited on f-quartz and c-sapphire substrates at different temperatures; (b) absorbance spectra for the same films at short wavelengths. The films were deposited at $I_{proc}/I_{met} \approx 0.5$.

The sharp increase in absorptance below 300 nm is due to the onset of the fundamental absorption of $Ga_2O_3$. The optical band gap ($E_g$) of ~5.0–5.1 eV for both the crystalline and X-ray amorphous films was estimated by a Tauc plot assuming direct band gap transitions (Fig. S6(a)), and is in good agreement with values obtained by applying the same procedure to magnetron sputtered films [29,34,35]. The changes in the phase and structure of the film can often be indirectly evaluated by observing a fundamental absorption edge shift. A slight increase in $E_g$ can be observed for both types of substrates when the deposition temperature is increased (Fig. S6(b)). The band gap of $Ga_2O_3$ can be affected by deviation from stoichiometry, i.e., excess gallium or a deficiency of oxygen [1], or by anisotropy of crystalline orientation [36].

**Conclusions**

We have demonstrated the feasibility of depositing stoichiometric $Ga_2O_3$ thin films by reactive pulsed-DCMS from a liquid Ga metal target. The practical problem with metallic Ga, in which it contracts into islands leaving parts of the target container uncovered and exposed to sputtering, was solved by pre-coating the stainless steel container by a wettability-promoting layer of carbon. No difficulties were encountered when conducting the sputtering process. The deposition rates of up to 37 nm/min at room temperature on f-quartz and 5 nm/min at 800°C on c-sapphire were higher than the data in the literature for RF sputtering of 22 and 1 nm/min, respectively. In line with the results in the literature for RF sputtering, the deposition rate was found to depend on the temperature of the substrate, with the atom's probability of sticking to the substrate decreasing significantly with temperature, and the factor limiting the deposition rate shifting gradually from the rate of arrival to the probability of sticking of the atoms to the surface as the substrate temperature was increased from RT to 800°C. We believe that there is still room for a further increase in the deposition rate, for instance by increasing the sputtering power and optimizing the process parameters.

At substrate temperatures ranging from RT to 500°C, a feature of the X-ray diffractograms is observed that may be indicative of the presence of small crystallites of β-$Ga_2O_3$ in the predominantly amorphous films. Pronounced XRD maxima start to appear at 500°C. Under the same sputtering conditions, the crystal structure depends on the substrate. For both f-quartz and c-sapphire, this is the β-$Ga_2O_3$ phase, but whereas differently oriented crystallites are detected for f-quartz, epitaxial films with a single orientation grow on c-sapphire above 600°C. These films are dense, with the surface becoming smoother as the substrate temperature increases. The dependence of the structure on the temperature is consistent with the data in the literature for other methods of deposition. Our results also indicate that post-annealing may be an alternative to heating the substrate during deposition to obtain crystalline films.

The films exhibit a low absorbance of less than 1% in the visible range. There is a slight blue-shift of the fundamental absorption edge with an increase in the substrate temperature, apparently because there are fewer defects in the films. No differences in the optical band gap (∼5.0–5.1 eV, estimated from Tauc plots) were observed between the films on f-quartz and c-sapphire. The optical properties are consistent with the data in the literature for other deposition methods.

**Declaration of competing interest**

The authors declare the following financial interests/personal relationships which may be considered as potential competing interests: A. Azens, M. Zubkins, E. Butanovs, and J. Purans (applicant to the Institute of Solid State Physics University of Latvia) have a national patent pending (No. LVP2021000105) and a European patent (No. EP22195507.3) pending.

**Acknowledgements**

This study was financially supported via ERDF project No. 1.1.1.1/20/A/057 "Functional ultrawide bandgap gallium oxide and zinc gallate thin films and novel deposition technologies". The Institute of Solid State Physics, University of Latvia, as a Center of Excellence, has received funding from the European Union's Horizon 2020 Framework Programme H2020-WIDESPREAD-01-2016-2017-TeamingPhase2 under grant agreement No. 739508, project CAMART². Support for the ERDA measurements from the Ion Technology Centre (ITC) at Uppsala University is gratefully acknowledged.

**References**

[1] Y. Yuan, W. Hao, W. Mu, Z. Wang, X. Chen, Q. Liu, G. Xu, C. Wang, H. Zhou, Y. Zou, X. Zhao, Z. Jia, J. Ye, J. Zhang, S. Long, X. Tao, R. Zhang, and Y. Hao, Fundam. Res. 1, 697 (2021).

[2] J. Shi, J. Zhang, L. Yang, M. Qu, D. C. Qi, and K. H. Zhang, Adv. Mater. 33, 2006230 (2021).

[3] S. J. Pearton, J. Yang, P. H. Cary, F. Ren, J. Kim, M. J. Tadjer, and M. A. Mastro, Appl. Phys. Rev. 5, 011301 (2018).

[4] M. Razeghi, J. Park, R. McClintock, D. Pavlidis, F. H. Teherani, D. J. Rogers, B. A. Magill, G. A. Khodaparast, Y. Xu, J. Wu, and V. P. Dravid, Proc. SPIE 10533, 105330R (2018).

[5] E. Ahmadi and Y. Oshima, J. Appl. Phys. 126, 160901 (2019).

[6] M. Bosi, P. Mazzolini, L. Seravalli, and R. Fornari, J. Mater. Chem. C 8, 10975 (2020).

[7] D. Yang, B. Kim, T. H. Eom, Y. Park, and H. W. Jang, Electron. Mater. Lett. 18, 113 (2022).

[8] J. Kim, T. Sekiya, N. Miyokawa, N. Watanabe, K. Kimoto, K. Ide, Y. Toda, S. Ueda, N. Ohashi, H. Hiramatsu, H. Hosono and T. Kamiya, NPG Asia Mater. 9, e359 (2017).

[9] S. Li, C. Yang, J. Zhang, L. Dong, C. Cai, H. Liang, and W. Liu, Nanomaterials 10, 1760 (2020).

[10] H. Liang, S. Cui, R. Su, P. Guan, Y. He, L. Yang, L. Chen, Y. Zhang, Z. Mei, and X. Du, ACS Photonics 6, 351 (2019).

[11] Z. Chi, J. J. Asher, M. R. Jennings, E. Chikoidze, and A. Pérez-Tomás, Materials 15, 1164 (2022).

[12] S. J. Pearton, F. Ren, M. Tadjer, and J. Kim, J. Appl. Phys. 124, 220901 (2018).

[13] M. Higashiwaki and G. H. Jessen, Appl. Phys. Lett. 112, 060401 (2018).

[14] K. Arora, N. Goel, M. Kumar, and M. Kumar, ACS Photonics 5, 2391 (2018).

[15] A. Pérez-Tomás, E. Chikoidze, Y. Dumont, M. R. Jennings, S. O. Russell, P. Vales-Castro, G. Catalan, M. Lira-Cantú, C. Ton–That, F. H. Teherani, V. E. Sandana, P. Bove, and D. J. Rogers, Mater. Today Energy 14, 100350 (2019).

[16] G. Seryogin, F. Alema, N. Valente, H. Fu, E. Steinbrunner, A. T. Neal, S. Mou, A. Fine, and A. Osinsky, Appl. Phys. Lett. 117, 262101 (2020).

[17] H. Hu, C. Wu, N. Zhao, Z. Zhu, P. Li, S. Wang, W. Tang, and D. Guo, Phys. Status Solidi 218, 2100076 (2021).

[18] K. Sasaki, A. Kuramata, T. Masui, E. G. Villora, K. Shimamura, and S. Yamakoshi, Appl. Phys. Express 5, 035502 (2012).

[19] D. J. Comstock and J. W. Elam, Chem. Mater. 24, 4011 (2012).

[20] S. Khartsev, N. Nordell, M. Hammar, J. Purans, and A. Hallén, Phys. Status Solidi (B) 258, 2 (2021).

[21] A. K. Saikumar, S. D. Nehate, and K. B. Sundaram, ECS J. Solid State Sci. Technol. 8, Q3064 (2019).

[22] W. D. Sproul, Vacuum 51, 641 (1998).

[23] P. J. Kelly and R. D. Arnell, Vacuum 56, 159 (2000).

[24] J. T. Gudmundsson, Plasma Sources Sci. and Technol. 29, 113001 (2020).

[25] A. Prabaswara, J. Birch, M. Junaid, E. A. Serban, L. Hultman, and C. L. Hsiao, Appl. Sci. 10, 3050 (2020).

[26] Y. Zhang, H. J. Whitlow, T. Winzell, I. F. Bubb, T. Sajavaara, K. Arstila, and J. Keinonen, Nucl. Instr. Phys. Res. B 149, 477 (1999).

[27] C. V. Ramana, E. J. Rubio, C. D. Barraza, A. Miranda Gallardo, S. McPeak, S. Kotru, and J. T. Grant, J. Appl. Phys. 115, 043508 (2014).

[28] W. Cui, Q. Ren, Y. S. Zhi, X. L. Zhao, Z. P. Wu, P. G. Li, and W. H. Tang, J. Nanosci. Nanotechnol. 18, 3613 (2018).

[29] S. Li, S. Jiao, D. Wang, S. Gao, and J. Wang, J. Alloys Compd. 753, 186 (2018).

[30] M. Ogita, K. Higo, Y. Nakanishi, and Y. Hatanaka, Appl. Surf. Sci. 175–176, 721 (2001).

[31] K. H. Choi and H. C. Kang, Mater. Lett. 123, 160 (2014).

[32] P. Marie, X. Portier, and J. Cardin, Phys. Status Solidi (A) 205, 1943 (2008).

[33] J. Ahman, G. Svensson, J. Albertsson, Acta Crystallogr., Sect. C: Cryst. Struct. Commun. 52, 1336 (1996).

[34] J. Wang, L. Ye, X. Wang, H. Zhang, L. Li, C. Kong, and W. Li, J. Alloys Compd. 803, 9 (2019).

[35] Y. Liao, S. Jiao, S. Li, J. Wang, D. Wang, S. Gao, Q. Yu, and H. Li, Cryst. Eng. Comm. 20, 133 (2018).

[36] N. Ueda, H. Hosono, R. Waseda, and H. Kawazoe, Appl. Phys. Lett. 71, 933 (1997).